\documentstyle[11pt,newpasp,twoside,epsf]{article}
\def\edcomment#1{\iffalse\marginpar{\raggedright\sl#1\/}\else\relax\fi}
\reversemarginpar

\begin{document}
\title{Small-scale Substructure in Dark Matter Haloes:\\
Where Does Galaxy Formation Come to an End?}
 \author{J.~E.~Taylor (1)} 
 \author{J.~Silk (1) \& A.~Babul (2)}
 \affil{(1) Astrophysics, University of Oxford\\ 
Denys Wilkinson Building, Keble Road, Oxford, OX1 3RH, UK}
 \affil{(2) Physics and Astronomy, University of Victoria\\ 
Elliott Building 3800 Finnerty Road, Victoria, BC, V8P 1A1, Canada }
\begin{abstract}
Models of structure formation based on cold dark matter predict that
most of the small dark matter haloes that first formed at high redshift 
would have merged into larger systems by the present epoch. Substructure 
in present-day haloes preserves the remains of these ancient systems, 
providing the only direct information we may ever have about the low-mass 
end of the power spectrum. I describe some recent attempts to model halo 
substructure down to very small masses, using a semi-analytic model of 
halo formation. I make a preliminary comparison between the model predictions,
observations of substructure in lensed systems, and the properties 
of local satellite galaxies.
\end{abstract}
%%%%%%%%%%%%%%%%%%%%%%%%%%%%%%%%%%%%%%%%%%%%%%%%%%%%%%%%%%%%%%%%%%%%%%%%%%
\section{Introduction}

This goal of this panel session was to assess the problems faced by 
cold dark matter (CDM). I will discuss one of these in particular, 
the apparent conflict between theory and observations on sub-galactic scales, 
known variously as the substructure problem or the dwarf galaxy problem. 
In the standard CDM picture of galaxy formation, dark matter haloes are
the sites for gas cooling and star formation, so there should be a close 
correspondence between luminous galaxies and individual dark matter haloes
or `subhaloes', the bound remnants of haloes that have merged into larger 
virialised systems. Theory and simulations predict that by $z=0$ the halo 
and subhalo mass functions should be relatively steep power laws on mass 
scales smaller than galaxy clusters. Galaxies, on the other hand, appear to 
have a limited range of luminosities and masses. The galaxy luminosity 
function can only be sampled down to its faintest limit in the very nearby 
universe, but it generally appears to be flat or slowly rising. Assuming a 
monotonic relation between average galaxy luminosity and halo mass, this 
leaves two possibilities; either CDM models are incorrect and there is 
little small-scale structure in the universe, or the net efficiency of 
galaxy formation (in the broadest sense of the amount of stellar light 
produced within an individual halo for a given halo mass) must decrease 
dramatically on small scales. 

There is also increasing observational evidence that the faint-end slope 
of the galaxy luminosity function varies systematically with environment 
(e.g.\ the talk by Tully and the poster by Roberts in these proceedings). 
Whatever mechanism suppresses the formation of dwarf galaxies, it must 
also reproduce this environmental dependence. Of course the dependence may 
not be primordial; effects such as galaxy harassment and ram-pressure 
stripping could have modified the luminosity function differently in 
different environments. The trend is the reverse of what one would naively
expect, however - there are more dwarf galaxies per giant galaxy in 
high-density environments, where harassment and stripping might be expected 
to reduce their relative number.

On large scales there is now strong observational support for a 
cosmological model whose matter density is dominated by a dark, 
non-baryonic component (Spergel, these proceedings).
It would be possible to reproduce most of these observations with variants 
of CDM in which structure is reduced on small scales, such as
warm, annihilating or collisional dark matter, or CDM with
a truncated power spectrum; thus resolving substructure problem. 
There are tentative indications from recent lensing observations that 
dark substructure has now been detected, however, so modifying CDM may 
be the wrong solution to the problem. In what follows I will assume the 
CDM picture of structure formation is correct on all scales, and discuss 
what the small-scale distribution of dark matter can then tell us about 
galaxy formation.

\section{A Brief History of the Dwarf Galaxy Problem}

The discrepancy between the field luminosity function and the 
predictions of hierarchical models was noted before the modern 
conception of CDM itself (White and Rees 1978). Kauffmann, White 
and Guiderdoni (1993) made the more specific observation that 
satellite galaxies should far outnumber the dozen dwarf companions 
seen around each giant galaxy in the Local Group. They proposed 
several possible solutions to the problem, but could not conclude 
definitively on the role of dynamics in disrupting dwarfs. 
It took high-resolution numerical simulations (Klypin et al.\ 1999; 
Moore et al.\ 1999) to prove convincingly that large numbers of 
subhaloes would survive in a system like the Milky Way.

In their analysis, Moore et al.\ (1999) assumed that the observed dwarfs 
trace out the full depth of their surrounding dark matter potential, and 
thus that the velocity dispersion of their stellar component is 
roughly $1/\sqrt{2}$ times the peak circular velocity of the corresponding 
dark matter structure. This led to an overall discrepancy of a
factor of 50 between the number of dwarfs and the number of subhaloes
that could host them. Subsequently, White (2000) pointed out that the 
stars in observed dwarfs might be condensed systems within much 
larger potential wells. Detailed modelling (Stoehr et al.\ 2002; 
Hayashi et al.\ 2003) confirmed that the satellites of the Milky Way 
could in fact reside in the most massive subsystems within its halo 
without this being apparent from their velocity dispersion profiles.

These two different pictures of the suppression of galaxy formation 
on small scales are illustrated schematically in Fig.\ 1. In the first 
case (left-hand panel), 
visible satellites are rare objects (only one subhalo in 100 
on the smallest scales), but the masses inferred from rotation curves 
(red line) are close to the total mass for each system (black curve). 
In the second case (right-hand panel), dwarfs form with widely varying 
efficiency in the dozen most massive subsystems, and galaxy formation stops 
altogether below some fairly high mass cutoff. These two scenarios imply
very different underlying regulatory mechanisms; a simple one 
depending only on the mass or circular velocity of halo 
(e.g.\ Dekel \& Silk 1986; Stoehr et al.\ 2002) in the case of 
the latter model ,
or one based on some other property of the system such as its age
(Bullock, Kravtsov, \& Weinberg 2000; Benson et al.\ 2002) 
in the case of the former.
Frustratingly, even very 
detailed observations of dwarf luminosities and kinematics may not be 
able to distinguish between these possibilities (Stoehr et al.\ 2002). 
Additional information is needed.

\begin{figure}
\plottwo{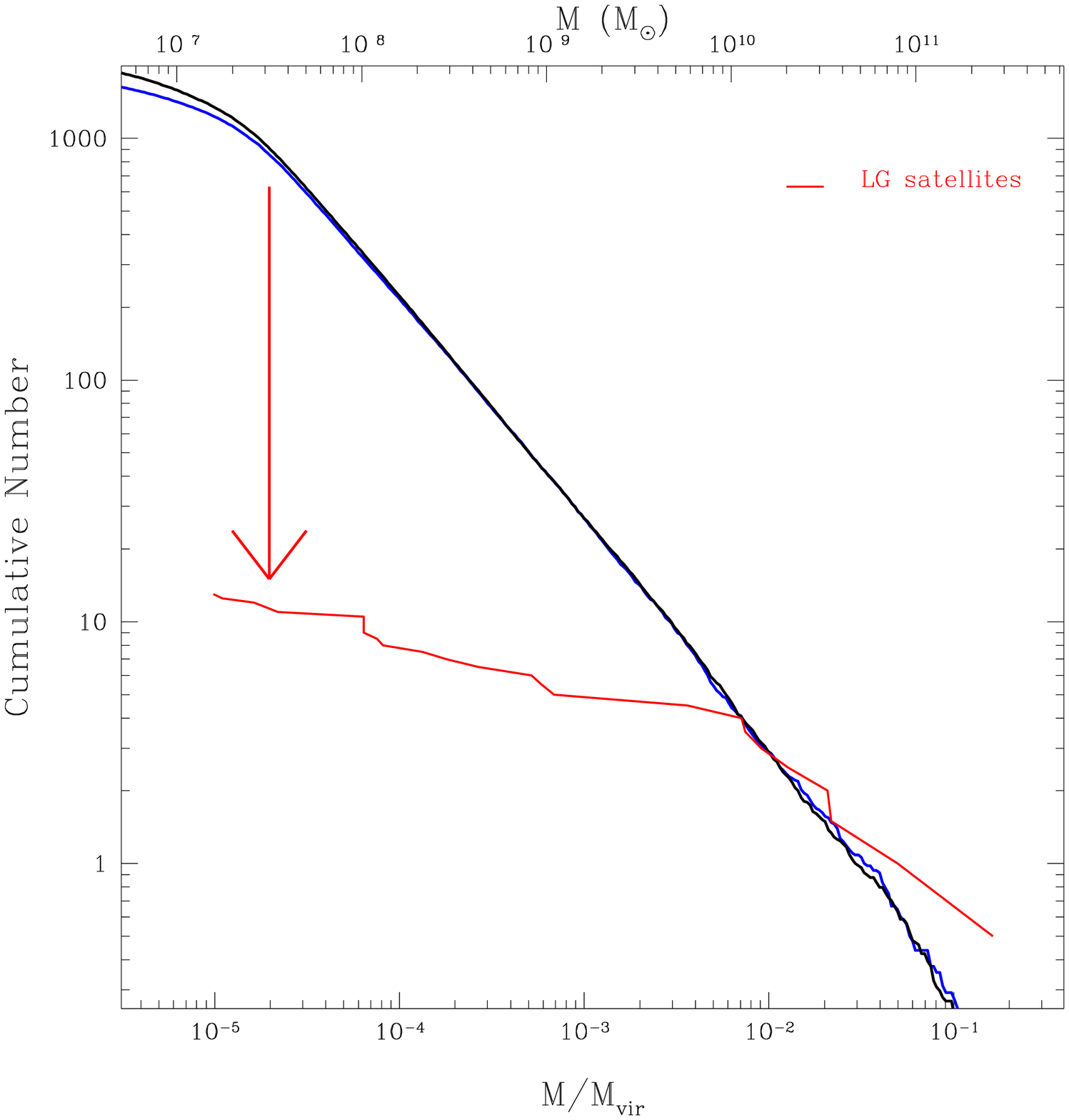}{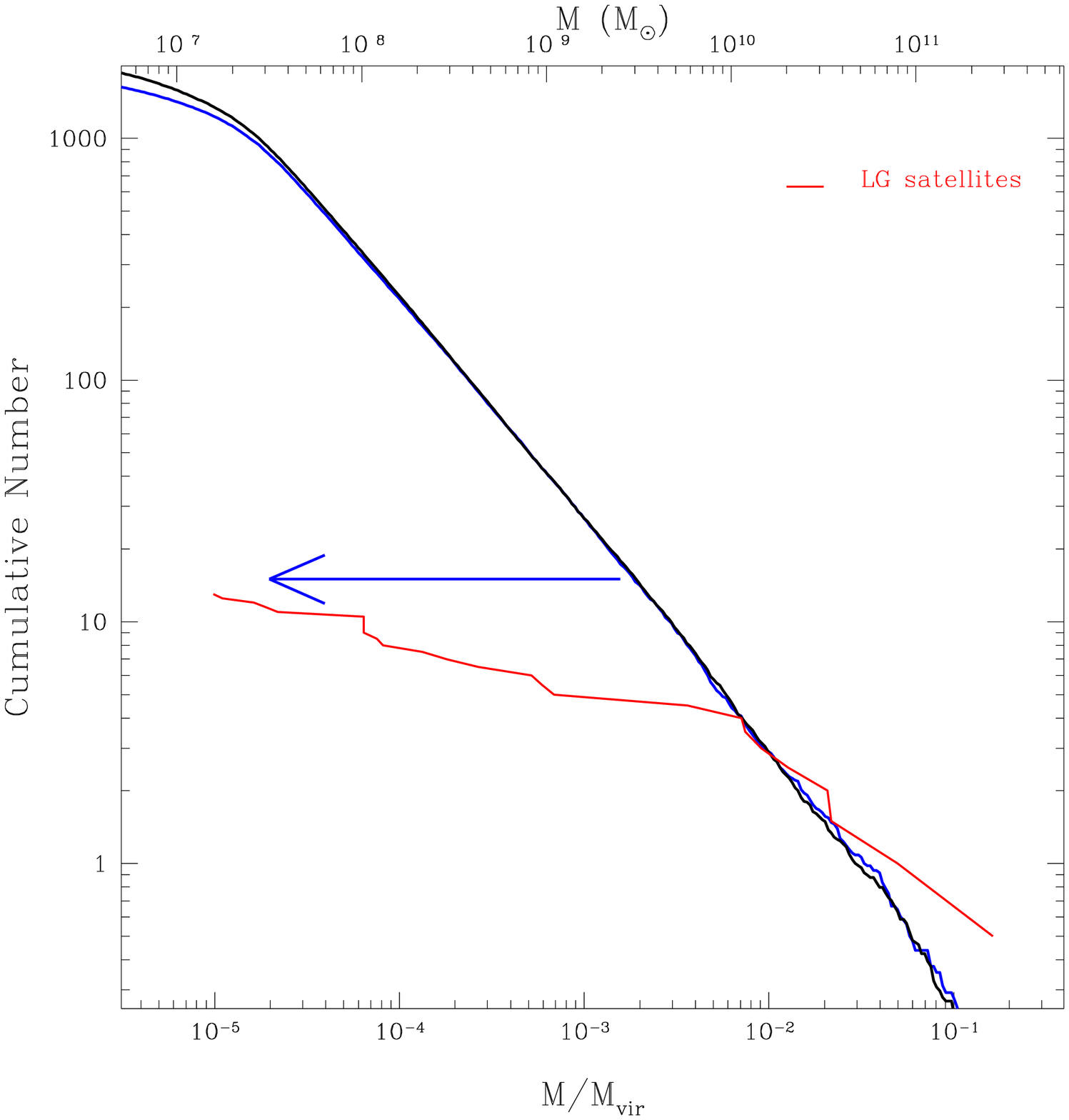}
\caption{Two models for the suppression of dwarf galaxies.
Model 1 (left-hand panel): visible dwarfs are rare but form over
a wide range of halo mass; Model 2 (right-hand panel): dwarf galaxies 
form exclusively in the most massive haloes.
}
\end{figure}

\section{Modelling Substructure: Do We Know Overmerging is Over?}

Given the complex non-linear dynamics involved in structure formation,
self-consistent numerical simulations are the preferred method 
for studying CDM halo formation and halo substructure. Simulations suffer 
from one serious drawback, however -- their results are least reliable in 
the densest parts of the halo, where
the dynamical timescale is shortest and artificial heating 
has the greatest effect. The artificial smoothing of the density 
distribution in these regions is referred to as `overmerging'
(for a review of the problem see Moore 2000). 

In early simulations of the formation of galaxy clusters, overmerging 
erased substructure completely (e.g.\ White 1976). When simulations reached 
sufficient resolution to resolve roughly as many subsystems as 
there are galaxies in a cluster, the problem was considered 
`solved' (e.g.\ Ghigna et al.\ 2000), although the scale invariance 
of halo properties quickly lead to an excess dwarf satellite problem 
in galaxy haloes. Interestingly enough, the assumption that the overmerging 
problem is now solved has not been fully tested. At current mass and force
resolution, simulations converge on the properties of massive
subhaloes in the outermost regions of a halo (e.g.\ Stoehr 2002).
Convergence tests in the innermost regions, where overmerging
should be strongest, have not been presented. These regions contain
only a small fraction of the total substructure in any given mass range,
so central overmerging would not be apparent in convergence tests
averaged over the whole halo; for applications such as lensing,
however, central substructure is crucially important. 

As an independent test of overmerging, we have compared the substructure 
in several high-resolution simulations of CDM haloes with the subhalo 
populations predicted by a semi-analytic model 
of halo formation (Taylor \& Babul 2001, 2003; Taylor 2001). 
Fig.\ 2 compares the spatial distribution of subhaloes
in the semi-analytic (upper solid lines) and numerical (lower lines 
with error bars). In the outer regions, the two techniques agree on the
normalisation of the halo mass function. In the inner regions,
the semi-analytic model, which should not be affected by overmerging,
predicts up to ten times more subhaloes. It is not clear which of these
results is correct -- the disruption rate in the semi-analytic model
could be artificially low. Increasing the resolution of the simulations 
(three panels, from right to left) increases the amount of central 
substructure, however, with no sign of convergence in the simulations 
presented here. This suggests that the simulations probably still 
underestimate the amount of central substructure in haloes, 
while the semi-analytic 
model may provide a more reliable estimate.

\begin{figure}
\begin{center}
\plotfiddle{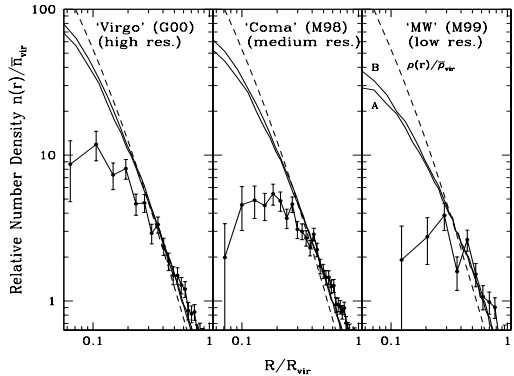}{4.5cm}{0}{35}{30}{-150}{-50}
\caption{The number density of subhaloes in three simulations of 
increasing resolution from right to left (jagged lines with error bars 
-- from Moore et al. 1998, 1999; Ghigna et al.\ 2000), compared with 
the semi-analytic prediction (smooth solid lines) and the halo density 
profile (dashed line). Each quantity is normalised to its mean value 
within the virial radius.
}
\end{center}
\end{figure}

The existence of small-scale structure can be confirmed observationally by
detecting its gravitational effects in lensing systems. Small structures 
in the halo of a lens can produce flux ratios between pairs or sets of 
images that cannot be explained by a smooth potential (Schneider, these 
proceedings). The analysis of several systems has already produced evidence 
for halo substructure amounting to 1--2\% of the projected mass density
of the inner parts of lensing systems (Dalal \& Kochanek 2002), 
but these estimates may be affected by stellar microlensing or other 
complicating effects. Future observations of the bending of background 
radio jets (Metcalf 2002) or spectrally resolved flux anomalies 
(Metcalf et al.\ 2003) will clarify the situation. 

\section{Satellites vs. Subhaloes}

I will now discuss the average properties of substructure in a set of 
144 semi-analytic haloes, generated in a Lambda-CDM cosmology with 
parameters consistent with recent WMAP results (Spergel, these proceedings). 
The haloes have a final mass of 
$1.6\times10^{12} M_{\odot}$ and their merger histories are resolved
down to $5\times10^{7} M_{\odot}$ (given the typical mass loss of 
subhaloes, this means the final substructure mass function is mostly 
complete down to a few million solar masses) and back to $z = 30$. 
I will consider whether dwarf galaxies can be associated with the 
subhaloes as a function of their original mass (before tidal stripping 
in larger systems), their original circular velocity, or their age.

\subsection{The Clustering of Satellites}

\begin{figure}
\plottwo{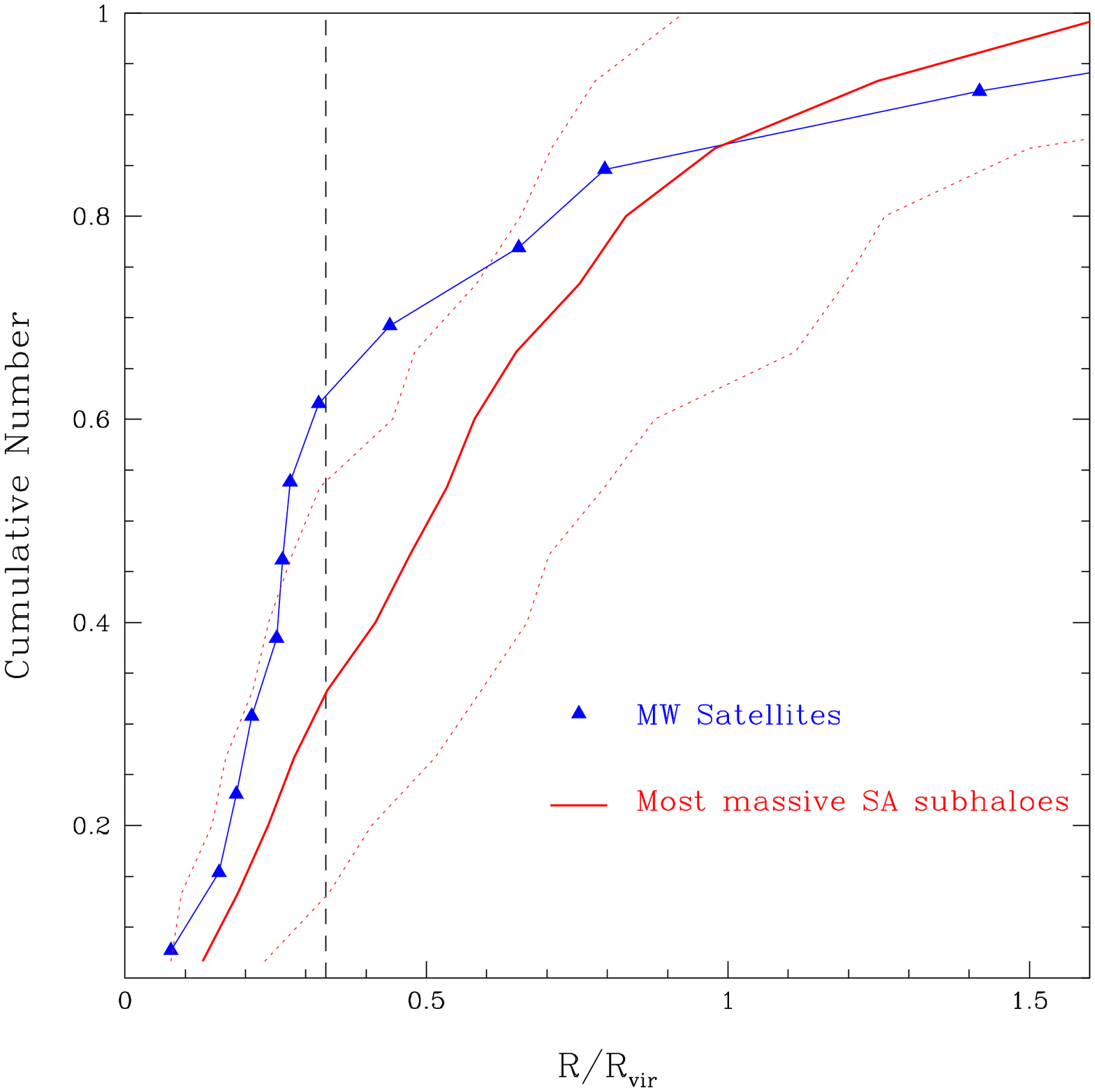}{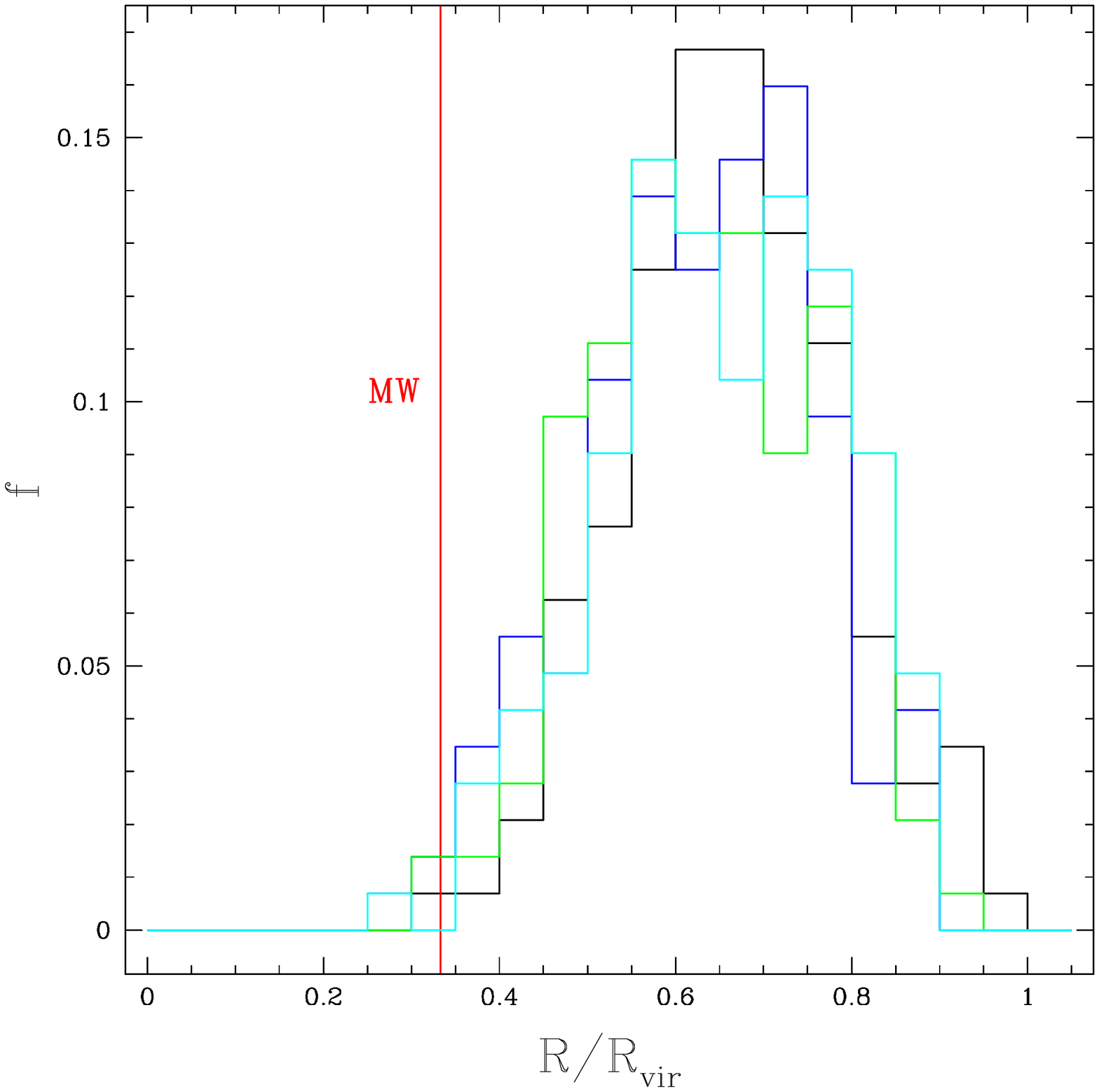}
\caption{(Left-hand panel) The cumulative number of the dozen 
most massive subhaloes
within some fraction of the virial radius. The solid line shows
the average distribution for all 144 models, while the dotted lines show
the 10\% and 90\% contours of the distribution. The blue
line and triangles show the distribution of Milky Way satellites.
(Right-hand panel) Histograms of the fraction of the virial
radius containing 8 of the 11 most massive subhaloes, for
each of the 144 model haloes. The different colours are for
different model parameters; the vertical line shows the value
for the Milky Way.
}
\end{figure}

Fig.\ 3 shows the cumulative radial distribution of the dozen 
subhaloes with the greatest original masses, normalised to the
present-day virial radius of the halo, and averaged over the
full set of trees. The dotted lines indicate the 10\% and 90\% 
contours of the distribution. The solid line and triangles show 
the distribution of satellites for the Milky Way, assuming a virial 
radius of 314 kpc, corresponding to a mass of $1.6\times 10^{12} M_{\odot}$.
If this normalisation is correct, then the satellites of the Milky Way 
are clearly more clustered than the most massive subhaloes in its
halo should be.

To illustrate the statistical strength of this conclusion more clearly,
Fig.\ 4 shows histograms of the normalised radius containing 
the innermost 8 of the 11 most massive satellites, for all 144 model haloes. 
The different line colours show results for different values for the 
Coulomb logarithm, the mass-loss parameter or the mean orbital
circularity in the semi-analytic model (see Taylor \& Babul 2003). The vertical
line indicates the value for the Milky Way. No more than 1--2 of the 
144 trees (depending on the model assumed) are as clustered as the satellites
of the Milky Way. Thus, the visible satellites of the Milky Way are more 
clustered than its most massive subhaloes at a $\sim 3$-$\sigma$ level. 

\subsection{Satellite Kinematics}

There is at least one important complication in the preceding
analysis. The halo of the Milky Way could be less massive than assumed, 
and therefore have a smaller virial radius and a less extended
population of satellites. In this case, the normalisation of the
radii used in Figs. 3 \& 4 would be incorrect. Reducing the virial
radius of the Milky Way by 20--30\% would bring the observed distribution
into perfect agreement with the predictions for massive subhaloes; 
this would imply a total halo mass of $\sim 8\times 10^{11} M_{\odot}$, 
which is still within the range of estimates
for the Milky Way (e.g.\ Klypin, Zhao, \& Somerville 2002). 

\begin{figure}
\plottwo{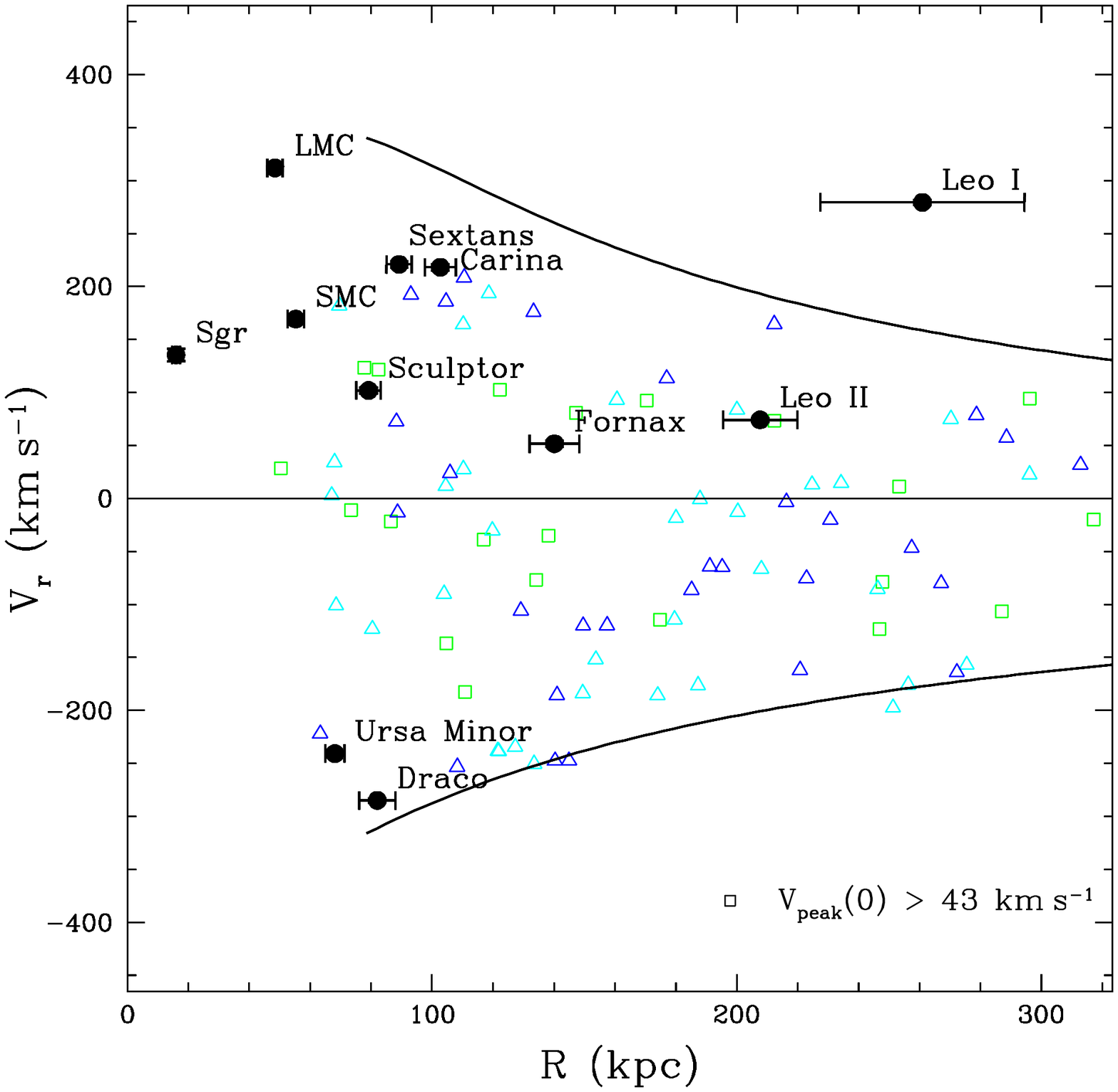}{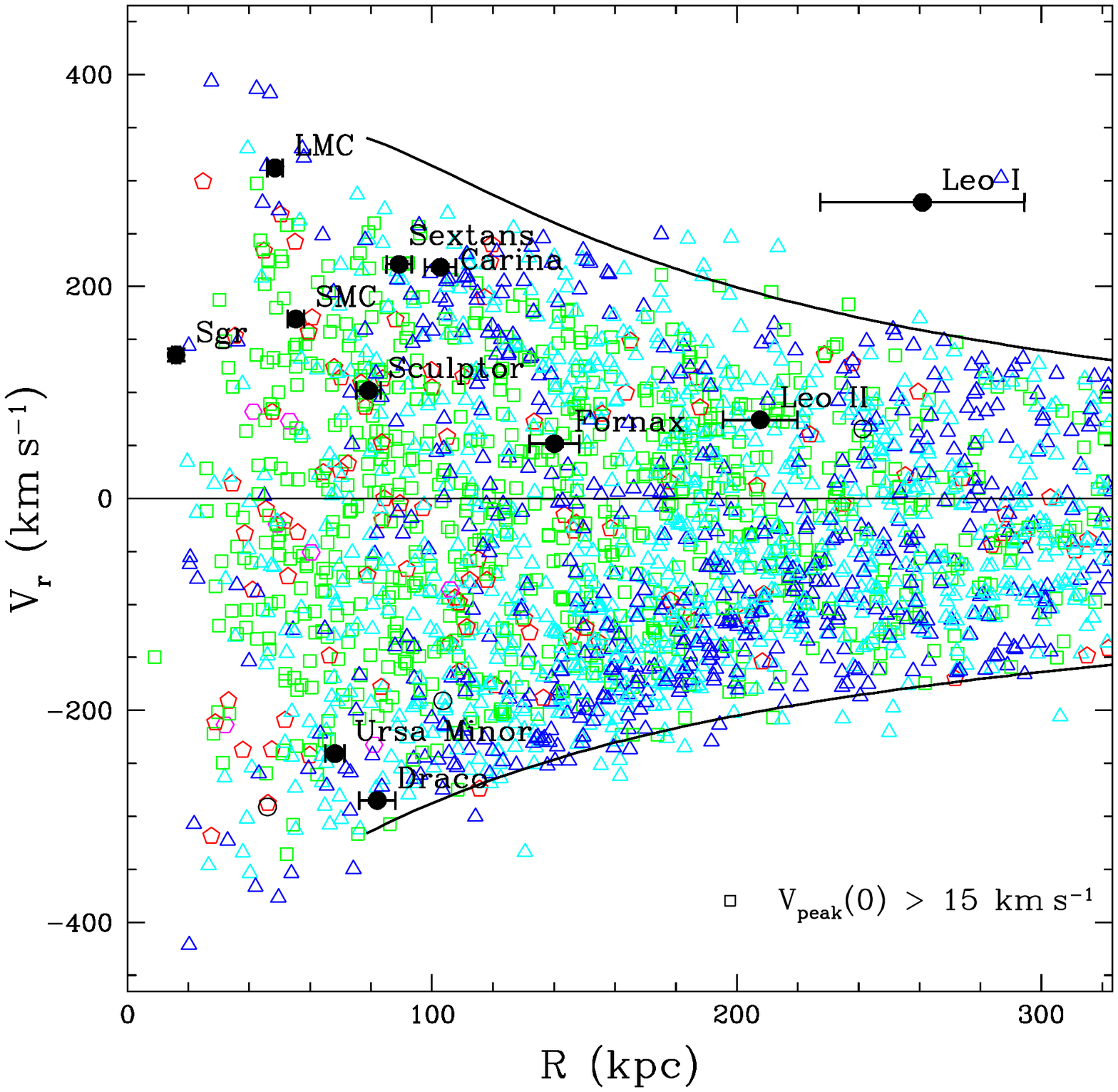}
\caption{Positions and radial velocities of Milky Way satellites
(solid circles with error bars), compared the distribution of
massive subhaloes (left-hand panel; open symbols) and low-mass 
subhaloes (right-hand panel). The solid lines indicate the envelope 
of the distribution, and the symbol types indicate the merger epoch 
(see text).
}
\end{figure}

This possibility can be tested by looking at satellite kinematics. The local
flow around the halo of the Milky Way is quite quiet; its satellites
have large radial velocities because they have been accelerated by its
potential. If the mass of the halo is smaller, the satellites will have
lower radial velocities. Fig.\ 5 shows radial velocity versus radius,
for the luminous satellites of the Milk Way (large filled circles 
with error bars) and for all the subhaloes over some (original) circular 
velocity limit (open symbols), selected from each of a small set of 
semi-analytic haloes. 
The symbol types indicate the merger epoch $z_{\rm m}$ of the subhalo;
triangles for $z_{\rm m} < 2$, squares for $z_{\rm m} = $2--4, pentagons for 
$z_{\rm m} = $4--6, hexagons for $z_{\rm m} = $6--8, 
and circles for $z_{\rm m} > 8$.
The thick lines give a rough envelope for the distribution of subhaloes.

The left-hand panel shows the results for massive systems. 
Several points are worth noting. First, the observed satellites are 
concentrated at smaller radii than the most massive subhaloes 
(left-hand plot), as discussed previously. On the other
hand, the observed scatter in radial velocity is consistent with the
model, which assumes a total halo mass of $1.6\times\,10^{12} M_{\odot}$. 
Reducing this by a factor of 2 would reduce the velocity dispersion by 
20-30\%, which is incompatible with the observed velocity of several 
satellites, notably the LMC. In fact even assuming this mass for the halo, 
there is no object like Leo I among the most massive subhaloes. 
Leo I has such a large radial velocity away from the Milky Way that it
must have passed through the potential of its halo on a very
radial orbit. Even in this case, however, if it had been massive it would
have lost a substantial amount of orbital energy through dynamical friction.
Thus Leo I cannot be a massive system.

The right-hand panel shows the same comparison for the less massive 
satellites. One again, the envelope matches the observed scatter
in radial velocities, suggesting the assumed mass of the halo is roughly
correct. Now there are a few systems like Leo I; as before, they have
passed close to the centre of the halo and are heading out on very
radial orbits, but now their velocities are higher as they have
not experienced as much frictional drag. While the subhaloes are still
located at larger radii than the luminous satellites on average, there 
is a strong trend in the distribution with merger epoch. Selecting systems 
which merged into the halo before a redshift of 4--6 (pentagons, hexagons 
or circles) produces the roughly the same degree of clustering as observed
for the luminous satellites. 

\section{Summary}

The predictions of large amounts of small-scale structure and 
substructure in CDM cosmologies is perhaps startling, but it is not
in and of itself a reason to reject CDM. Galaxy formation is unlikely
to be a simple process, particularly on small scales where many
regulatory mechanisms might limit its efficiency. Thus, if we fail
to see large numbers of dwarf galaxies in the nearby universe, this may
be telling us more about galaxy formation than it does about CDM. 
Assuming CDM substructure exists, its properties can provide
important constraints on the process of dwarf galaxy formation.

Simulations and semi-analytic models of halo substructure agree
on its properties in the outer parts of haloes; in the inner regions,
simulations may still suffer from overmerging, the artificial disruption of 
substructure due to numerical effects. Interestingly, the amount of central
substructure seen in the semi-analytic haloes considered here 
is consistent with the 
amount of substructure inferred from strong lensing experiments. 
Thus the semi-analytic haloes may provide a more accurate picture of the
spatial distribution of substructure around galaxies.

Comparing this distribution to the observed location of the Milky Way's 
luminous satellites, it is clear that observed satellites are more clustered
than the most massive subhaloes should be, at a $\sim 3 \sigma$ level. 
Satellite radial velocities strengthen this conclusion further, partly
by constraining the mass of the halo and partly through specific examples 
such as Leo I. This appears to rule out models of feedback where only the 
most massive subhaloes host dwarf galaxies, and favours models in which 
the oldest subhaloes do instead. More generally, it should also rule out 
many proposed alternatives to CDM, such as warm or interacting dark matter. 
Almost all of these scenarios would destroy the smallest subhaloes
and those in the centre of the main system preferentially, which
is the exact opposite of the pattern observed. 

\acknowledgments{I gratefully acknowledge support from the Leverhulme Trust 
during the course of this work.}

%%%%%%%%%%%%%%%%%%%%%%%%%%%%%%%%%%%%%%%%%%%%%%%%%%%%%%%%%%%%%%%%%%%%%%%%%%

%%%%%%%%%%%%%%%%%%%%%%%%%%%%%%%%%%%%%%%%%%%%%%%%%%%%%%%%%%%%%%%%%%%%%%%%%%

\begin{references}

\reference Benson A.~J., Lacey C.~G., Baugh C.~M., Cole S., Frenk C.~S., 2002, MNRAS, 333, 156 

\reference Bullock J.~S., Kravtsov A.~V., Weinberg D.~H., 2000, ApJ, 539, 517 

\reference Dalal N.~, Kochanek C.~S., 2002, ApJ, 572, 25 

\reference Dekel, A.~\& Silk, J.\ 1986, ApJ, 303, 39 

\reference Ghigna S., Moore B., Governato F., Lake G., Quinn T., Stadel J., 2000, ApJ, 544, 616 

\reference Hayashi E., Navarro J.~F., Taylor J.~E., Stadel J., \& Quinn T.\ 2003, ApJ, 584, 541 

\reference Kauffmann G., White S.D.M., Guiderdoni B., 1993, MNRAS, 264, 201

\reference Klypin A., Gottl{\" o}ber S., Kravtsov A.~V., Khokhlov A.~M., 1999, ApJ, 516, 530 

\reference Klypin A., Zhao H., Somerville R.~S., 2002, ApJ, 573, 597 

\reference Metcalf R.~B., 2002, ApJ, 580, 696 

\reference Metcalf R.~B., Moustakas L.~A., Bunker A.~J., Parry I.~R. 2003
preprint (astro-ph/0309738)


\reference Moore B.\ 2000, preprint (astro-ph/0009247)

\reference Moore B., Governato F., Quinn T., Stadel J., Lake G., 1998, ApJ, 499, L5

\reference Moore B., Ghigna S., Governato F., Lake G., Quinn T., Stadel J., Tozzi P., 1999, ApJ, 524, L19

\reference Stoehr F., White S.~D.~M., Tormen G., Springel V., 2002, MNRAS, 335, L84 

\reference Taylor J.~E., 2001, Ph.D. thesis (University of Victoria)

\reference Taylor J.~E., Babul A., 2001, ApJ, 559, 716

\reference Taylor J.~E., Babul A., 2003, MNRAS, submitted (astro-ph/0301612)

\reference White S.~D.~M.\ 1976, MNRAS, 177, 717 

\reference White S.~D.~M.\ 2000, ITP conference presentation, (online.itp.ucsb.edu/online/galaxy\_c00/white) 

\reference White S.~D.~M.~\& Rees, M.~J.\ 1978, MNRAS, 183, 341 



\end{references}
\end{document}